\documentstyle[12pt]{article}
\input epsf
\begin{document}
\begin{titlepage}
{\center
April, 1995 	\hfill		SU-ITP-95-9\\
		\hfill		CERN-TH/95-101\\
		\hfill		hep-ph/9504415\\}
\vskip .1in
\center{\Large\bf The Supersymmetric Flavor Problem}
\vskip .3in
\center{\sc Savas Dimopoulos$^a$\footnote{On leave of absence
from the Physics Department, Stanford University, Stanford, Ca.
94305, USA. Work supported in part by NSF Grant
NSF-PHY-9219345},David Sutter$^b$\footnote{Work supported in part by
NSF Grant
NSF-PHY-9219345 and an NSF Graduate Fellowship.
E-mail: sutter@squirrel.stanford.edu}}
\vskip .1in
{$^a$ {\it Cern Theory Group, 1211 Geneva 23, Switzerland}}\\
{$^b$ {\it Physics Department, Stanford University, Stanford, Ca.
94305, USA}}
\vskip .5in
\begin{abstract}
The supersymmetric $SU(3)\times SU(2)\times U(1)$ theory with
minimal
particle content and general soft supersymmetry breaking terms has
110 physical parameters in its flavor sector: 30 masses, 39 real
mixing angles and 41 phases.  The absence of an experimental
indication for the plethora of new parameters
places severe constraints on theories posessing Planck or GUT-mass
particles and suggests that theories of flavor conflict with
naturalness.
We illustrate the problem by studying the processes $\mu \rightarrow
e + \gamma$ and $K^0 - \bar{K}^0$ mixing  which are very sensitive
probes of Planckian physics: a single Planck mass particle coupled
to the electron or the muon with a Yukawa coupling comparable to the
 gauge coupling typically leads to a rate for $\mu \rightarrow e +
\gamma$ exceeding the present experimental limits. A possible
solution is that  the messengers which transmit supersymmetry
breaking to the ordinary particles are much lighter than $M_{\rm
Planck}$.
\end{abstract}

\end{titlepage}

\section{Naturalness versus Flavor: A Conflict}

Nature is ambivalent about Flavor; Quark masses violate it
significantly whereas neutral processes conserve it very accurately.
This ambivalence leads to a conflict that has to be resolved in
every
theory. In the standard model it led to the GIM mechanism. In
SUSY-GUTS and the supersymmetric standard model \cite{dg} it led to
the hypothesis that squarks and sleptons of the same color and
charge
have the same mass, independent of the generation that they belong
to. We call this ``horizontal universality''. A stronger version of
this hypothesis is that all squarks and sleptons have the same mass
at  $M_{\rm GUT}$ \cite{dg}. This is called ``universality'' and is
a
fundamental ingredient of the minimal version of the Supersymmetric
Standard Model (MSSM).  Universality ensures that the sparticle
masses are isotropic in flavor space and thus do not cause any
direct
flavor violations. Flavor non-conservation  in the MSSM originates
in
the quark masses and is under control.

The hypotheses of universality or horizontal universality are
difficult to implement in realistic theories. The reason is simple:
the physics that splits particles also splits sparticles. The degree
to which this happens is the crucial question; the answer is model
dependent. In the MSSM  and the minimal SUSY-GUT  \cite{dg} the
interfamily sparticle splittings that are dynamically induced are
adequately small. Such minimal theories leave fundamental questions
unanswered and are unlikely to be the last word. In more ambitious
theories addressing (even small parts of) the problem of flavor, the
interfamily sparticle splittings are invariably large and cause
unacceptable flavor violations unless the sparticle masses are heavy
\cite{hkr}.  However, heavy sparticles spoil naturalness, which was
the
original reason for low energy SUSY; it implies that parameters
related to electroweak symmetry breaking must be tuned to high
accuracy. Thus, generic theories addressing the problem of flavor
conflict with naturalness.  These theories require a new mechanism
to
supress flavor violations to solve this conflict.

In this paper, we first present the general supersymmetrized
standard
model.  Next we review sources of flavor dependance in the
supersymmetry violating terms and the naturalness criterion that
limits the masses of the new supersymmetric partners.  We then use
the experimental constraints from the flavor changing processes $\mu
\rightarrow e + \gamma$ and $K^0 - \bar{K}^0$ mixing to quantify the
conflict between sparticle non-universality and naturalness,
illustrating the need for a mechanism to supress flavor changing
processes.  Finally, we discuss a possible supression
mechanism.

\section{Supersymmetric Kobayashi-Maskawa.}

\subsection{The $U(3)^5$ Flavor Group}

Consider the general, renormalizable softly broken supersymmetric
$SU(3)\times SU(2)\times U(1)$ theory with minimal particle content
at some energy scale significantly smaller than the fundamental
scales $M_{\rm Planck}$ or $M_{\rm GUT}$. The soft SUSY-breaking
terms
\cite{dg,gg} are all taken to be $\sim M_{\rm weak}$ but are
otherwise
unconstrained: the sparticle masses are, in general, unrelated to
each other and the triscalar couplings are not necessarily
proportional to the Yukawa couplings\footnote{We also assume
R-parity
conservation.}.  This theory can be the low energy manifestation of
a
SUSY-GUT or a Superstring theory or anything else.  The gauge part
of
the Lagrangian has a $U(3)^5$ global {\it flavor} symmetry, one
$U(3)$ for each of the five species that constitute a family:
$q,\bar{d},\bar{u},l,\bar{e}$.  In this paper we study violations of
this flavor symmetry so we will concentrate on the flavor $U(3)^5$
breaking part of the Lagrangian, which is:
\begin{equation}
{\cal{L}}_{U(3)^5}^{break} = \sum_{A,i,j}{ m^{2}_{Aij}
\tilde{A}_{i}^{*} \tilde{A}_j }+ {\cal{L}}_{Yukawa} +
{\cal{L}}_{triscalar}
\end{equation}
\noindent where $\tilde{A}=\tilde{q}$, $\tilde{\bar{u}}$,
$\tilde{\bar{d}}$, $\tilde{l}$, $\tilde{\bar{e}}$ labels the five
species that constitute a family; the tilde labels a sparticle; and
i,j = 1,2,3 are U(3) flavor labels. The Yukawa part of the
Lagrangian
includes the corresponding scalar quartic couplings and is derived
from the superpotential:
\begin{equation}
{W}_{\rm Yukawa} = q{\lambda}_{u} \bar{u}H_u + q {\lambda}_{d}
\bar{d}H_d
+
l {\lambda}_{e}\bar{e}H_d
\end{equation}
\noindent where ${\lambda}_{u}, {\lambda}_{d}, {\lambda}_{e}$ are
Yukawa matrices.

The triscalar couplings are given by:
\begin{equation}
{\cal{L}}_{\rm triscalar} = \tilde{q} {A}_{u}  \tilde{\bar{u}}H_u +
\tilde{q}  {A}_{d} \tilde{\bar{d}}H_d +
\tilde{l} {A}_{e} \tilde{\bar{e}}H_d
\end{equation}
where $A_u, A_d, A_e$  are three by three Yukawa-like matrices with
overall magnitude of order $\sim M_{\rm weak}$.

\subsection{Counting Parameters}

The above Lagrangian contains three fermion Yukawa matrices, three
triscalar coupling matrices, and five scalar mass matrices.  The
Yukawa and triscalar  matrices are general $3 \times 3$ matrices
with
nine real magnitudes and nine imaginary phases each.  The five
scalar
mass matrices are $3 \times 3$ Hermitean matrices with six real
magnitudes and three phases each.  This gives a total of 84 real
parameters (mass eigenvalues and angles) and 69 phases.

Not all of these parameters are physical; some may be eliminated
using the $U(3)^5$ flavor symmetry of the gauge sector.  (We will
only allow superfield rotations in order to maintain the form of the
gaugino couplings)\footnote{Continuous R-Symmetry  will be used to
render the gaugino masses real.  Similiarly, Higgs fields  will be
redefined  to make the bilinear soft term real.}.  The $SU(3)^5$
subgroup of  $U(3)^5$ can be used to remove 15 real angles and 25
phases.  The $U(1)^5$ subgroup can be used to remove only three
phases.  The remaining two phases of  $U(1)^5$ can not be used to
remove any parameters because the Lagrangian is invariant under
these
rotations, which correspond to baryon and lepton number\footnote{In
doing a similiar counting for the standard model, the Lagrangian is
invariant under four $U(1)$ field redefinitions:  baryon number and
the three individual lepton numbers.}.

Subtracting the removeable parameters, we find the theory contains
69
real parameters and 41 phases. Of the 69 real parameters 30 are
masses, 9 for fermions and 21 for scalars, and the remaining 39 are
mixing angles. Thus, compared to the standard model, there are an\
additional 21 masses, 36 mixing angles and 40 phases. They all imply
new physics. A geometric interpretation of these parameters will
become clear in the next section.

\subsection{Sparticle Basis}
The first term of eq. 1 is quadratic and chirality conserving.  A
$U(3)^5$ rotation  can diagonalize it and take us to the
``sparticle'' basis \footnote{Unless otherwise specified we will
always make superfield rotations: sparticles and particles are
rotated in parallel. This ensures that the gaugino couplings have
their minimal form.} where:
\begin{equation}
 m^{2}_{Aij} =  m^{2}_{A(i)} \delta_{ij}
\end{equation}

Thus, in this basis, these chirality conserving terms of the
Lagrangian also conserve a $U(1)^{15}$  flavor subgroup that
conserves individual species number for each of the 15 species of
quarks and leptons that make up the three families. In the sparticle
basis, although the chirality conserving terms in the Lagrangian
distinguish the 15 species of sparticles, they do not cause flavor
violating transitions between them. This is convenient for tracing
flavor violations; they are associated with chirality violations and
originate either in the Yukawa superpotential or in the triscalar
couplings.

In this basis the Yukawa superpotential has the form:
\begin{equation}
{W}_{\rm Yukawa} = q U_{q} \bar{\lambda}_{u} U_{\bar{u}} \bar{u}
H_{u} +
q U_{q}' \bar{\lambda}_{d} U_{\bar{d}} \bar{d} H_{d} +
l U_{l} \bar{\lambda}_{e} U_{\bar{e}} \bar{e} H_{d}
\end{equation}
\noindent where $\bar{\lambda}_{u}$,$\bar{\lambda}_{d}$, and
$\bar{\lambda}_{e}$ are the diagonal Yukawa couplings for the quarks
and electrons.  $U_{q}'$, $U_{q}$, $U_{\bar{u}}$, $U_{\bar{d}}$,
$U_{l}$, and $U_{\bar{e}}$ are six unitary matrices;
$U_{q}^{\dagger}
U_{q}'$ is the usual KM matrix, whereas the remaining five are new
independent matrices.  In general, these matrices cannot be rotated
away. They have both physical and geometrical significance. Their
physical significance is that they cause new flavor violations.
Their
geometrical significance is that they measure the relative
misalignment between sparticle and particle masses in flavor
$U(3)^5$
space.  The A-terms are of a similiar form to the Yukawa couplings.
They contain six additional $3 \times 3$ unitary matrices with a
similiar physical interpretation.

\subsection{Universality and Proportionality}
In minimal supersymmetric theories it is often assumed that, at some
fundamental scale  $\sim M_{\rm GUT}$ or $M_{\rm string}$, each
triscalar
coupling is proportianal to the corresponding Yukawa coupling with a
proportionality constant which is the same for each Yukawa matrix.
This is sometimes called {\it proportionality} and reduces the
possible 27 complex numbers to one.  In addition, again in minimal
theories, one of two conditions is also postulated \cite{dg}:
\begin{itemize}
\item {{\bf horizontal universality}:
\begin{equation}
 m^{2}_{Aij} =  m^{2}_{A} \delta_{ij}
\end{equation}}
or, the more restrictive
\item {{\bf universality}:
\begin{equation}
 m^{2}_{Aij} =  m^{2} \delta_{ij}
\end{equation}}
\end{itemize}

Either version of universality reduces the sparticle masses to
spheres in flavor space which preserve the full $U(3)^5$ rotation
group. Since a sphere points nowhere, the notion of relative
orientation of particle and sparticle masses loses its meaning; the
geometric significance of 5 of the 6 matrices $U_{q}'$, $U_{q}$,
$U_{\bar{u}}$, $U_{\bar{d}}$, $U_{l}$, and $U_{\bar{e}}$
disappears. Only the usual CKM matrix $U_{q}^{\dagger} U_{q}'$ that
measures the relative orientation of up and down quark masses
continues to have geometrical and physical meaning. In particular,
since $U_{l}$ and $U_{\bar{e}}$ lose their meaning, there are no
lepton number violations in theories satisfying horizontal
universality.
The importance of the hypotheses of proportionality and universality
is now clear: They insure that all flavor violations involve the
quarks and are proportional to the usual CKM matrix $U_{q}^{\dagger}
U_{q}'$; consequently, they are under control.

As we shall review in the next section, the problem with these
hypotheses is that they do not seem to emerge from fundamental short
distance theories, such as GUTs or strings: Flavor breakings in the
fermion sector invariably pollute the soft terms and render them
non-universal and non-proportional.  This is, in one sense,
fortunate
because these low energy parameters may serve as a fingerprint of
high energy physics that is otherwise beyond the reach of
experiment.

Since we wish to do a general analysis of flavor violations we will
not assume proportionality or any form of universality.

\section{Sources of Non-Universality}

All theories have some degree of flavor dependence in the soft SUSY
breaking terms.  The terms which violate the $U(3)$ flavor symmetry
for the fermions will also affect the soft terms; if not at tree
level, then at the loop level through the RG equations.  The only
question is the extent to which the sparticles are non-degenerate
between families and misaligned with respect to the fermions.  The
answer depends on the size of the Yukawa couplings.

In theories that do not address the question of flavor, most of the
Yukawa couplings are small so they do not contribute significantly
to
flavor or CP violations; the top Yukawa is large, so it can induce
measurable violations \cite{bh}.  Recent calculations of $\mu
\rightarrow e + \gamma$, as well as electron and neutron electric
dipole moments,  in the minimal SUSY GUT  give results that could be
observed soon if sparticles are not too heavy \cite{bh}.  In this
case, the large top Yukawa does not cause a problem because it is
sheltered from the first generation by a small mixing angle.

In theories that do address the question of flavor \cite{an} we
expect that there are no small parameters and that  all
non-vanishing
Yukawa couplings are of the same order as the gauge coupling at some
high scale $ \sim  M_{\rm PL}$ (or $ \sim M_{\rm GUT}$), which we
call the flavor scale.  These Yukawas couple the three ordinary
families to superheavy multiplets residing at the flavor scale.  As
we shall demonstrate,  they can create large splittings among the
ordinary squarks and sleptons.  These subsequently lead to
dangerous
flavor violating interactions.  Even if there is a flavor symmetry
protecting the soft terms, threshold corrections will occur when the
symmetry is spontaneously broken, resulting once again in dangerous
contributions.

As an example, let us examine the $SU(5)$ superpotential term of
equation~\ref{eq:sup} which involves one light matter multiplet and
two heavy fields.  The RG equation describing the evolution of the
soft mass term for the matter multiplet is given in
equation~\ref{eq:ev}.
\begin{equation}
\label{eq:sup}
W = \lambda \bar{5}_{\rm family} \bar{5}_{\rm heavy} 10_{\rm heavy}
\end{equation}
\begin{equation}
\label{eq:ev}
\frac{dm_{\rm family}^2}{dt} = \frac{1}{8 \pi^2} 4 \lambda^2 (
m_{\rm family}^2 + m_{\bar{5} {\rm heavy}}^2 + m_{10 {\rm heavy}}^2
+ A^2)
\end{equation}

Let us first examine the running of $m_{\rm family}$ due to
evolution
from the string scale, $5 \times 10^{17}$ GeV, to the GUT scale, $2
\times 10^{16}$ GeV.  Assuming the Yukawa coupling is equal to one,
and the tri-linear mass $A$ is equal to the scalar masses $m$, we
obtain $\delta m^2= .65 m^2$ using the linear approximation.
Clearly
the linear approximation breaks down, but we do expect fractional
splittings of O(100\%) if there are large Yukawa couplings over a
broad range of energies.

If we redo the calculation, now assuming only a factor of two
between
the masses over which we integrate the RG equation instead of the
factor of 25 from the string scale to the GUT scale, the answer is
$\delta m^2=.14 m^2$.  This is a calculation typical of a threshold
correction estimate from a broken symmetry, and the answer is large,
as we shall see from flavor changing calculations.  Because there is
a logarithmic dependence on the ratio of mass scales, even a small
integration interval gives a significant mass correction.

In any theory of soft SUSY breaking terms, there must be a violation
of the flavor symmetry communicated through loop effects from the
flavor breaking Yukawa sector.  Small Yukawa couplings are not
dangerous, but in the presence of large Yukawa couplings, as
expected
near the flavor scale, there will be large flavor violations in the
SUSY breaking sector.

\section{Naturalness}

The naturalness criterion measures the sensitivity of the weak scale
to variations of the SUSY parameters at a fundamental scale, for
example the GUT scale.  In this section we will review a simplified
form of the analysis of Barbieri and Giudice \cite{bg}.

If the conditions for symmetry breaking are met, the minimum of the
Higgs potential at tree level can be written in terms of two
equations, one for $\tan{\beta}$ and one for $M_{\rm z}^{2}$.  The
latter
reads as:
\begin{equation}
\label{eq:mz}
M_{\rm z}^2 = 2 \frac{ (m_{Hd}^2 + \mu^2) -  (m_{Hu}^2 + \mu^2)
\tan{^2
\beta} }{ \tan{^2 \beta} - 1 }
\end{equation}
\noindent Here $m_{Hd}^2$ and $m_{Hu}^2$ are the soft scalar masses
of the down and up Higgs respectively, and $\mu$ is the Higgsino
mass.  All parameters in the above equation are evaluated at $M_z$.

The next step is to write this equation in terms of parameters at
the
GUT scale, for which one loop RG equations are sufficient.  For
clarity, we will keep $\tan{\beta}$, evaluated at the weak scale, in
the equation as a fundamental parameter.  This simplifies the
resulting equation, making $M_z^2$ linear in the GUT scale
parameters, without changing the numerical results significantly.
In
addition, we will keep the $\mu$ parameter evaluated at $M_{weak}$.
This does not effect the results because $\mu$ is renormalized only
by a multiplicative constant.

Equation~\ref{eq:mz2} gives $M_{\rm z}$ in terms of the parameters
of
interest.
\begin{eqnarray}
\label{eq:mz2}
M_{\rm z}^2 & = & c_{\mu} \mu^2 + c_{Hd} m_{Hd0}^{2} + c_{Hu}
m_{Hu0}^{2} +
c_{t}   m_{t0}^{2} + c_{\bar{t}} m_{\bar{t}0}^{2} +
\nonumber \\
& & c_{M} M_0^2 + c_{AM} A_{t0} M_0 + c_{A} A_{t0}^{2}
\end{eqnarray}

\noindent A subscript {\it 0} refers to a parameter evaluated at the
GUT scale\footnote{We will keep this convention throughout the
paper.}.  $M$ is the gaugino mass (unification is assumed), and
$m^2$
is a soft scalar mass.  The $c$ coefficients are functions of
$\tan{\beta}$ and constants of O(1) from RG solutions.  We have
assumed the top Yukawa is the only contributing Yukawa coupling.

There is no {\it a priori} relation among the $c$ coefficients, so
it
is unlikely that a large cancellation between seperate terms of
equation~\ref{eq:mz2} will occur.  We define the fine tuning of a
given term as the fraction by which $M_{\rm z}^2$ is smaller than
that
term.  For example, the fine tuning of the term associated with the
parameter $\mu$, which we label $f_{\mu}$, is given in
equation~\ref{eq:ftmu}.
\begin{equation}
\label{eq:ftmu}
f_{\mu} = \frac{M_{\rm z}^2}{c_{\mu} \mu^2}
\end{equation}
\noindent  Unless there is some cancellation mechanism, the limit to
a reasonable cancellation is usually placed at a fine tuning of
$f=.1$.  This is the 10\% naturalness criterion.

This analysis gives especially tight constraints on the parameters
$\mu$ and $M_0$.  Independent of $\tan{\beta}$ and the
renormalization group, the coefficient $c_{\mu}=2$.  The minimum
value of $c_{M}$  is   $\approx 6$ and occurs for $\tan{\beta} \gg
1$.  There is only a weak dependence on the size of the top Yukawa
because we are near the fixed point.  With these $c$ values, the
10\%
fine tuning criterion gives the following upper mass limits:
\begin{eqnarray}
\nonumber
M_0 = 117 GeV
\\
\nonumber
\mu = 203 GeV
\end{eqnarray}
\noindent  If one wants to allow for a larger fine tuning, the
square
of the masses can be scaled up by the factor by which the fine
tuning
is increased.  For example, a $1\%$ fine tuning gives an upper limit
on $M_0$ of 370~GeV.

\section{Flavor Changing Processes}

We will now calculate the constraints placed on sparticle
non-universality and SUSY masses by the processes $\mu \rightarrow e
+ \gamma$ and $K^0 - \bar{K}^0$ mixing.  To do this, we will make
two
simplifying approximations, valid in a large class of theories:  we
will neglect the contributions of the $A$ terms and the third
family.
Including these contributions will improve the bounds and will make
our results stronger.

\subsection{$\mu \rightarrow e + \gamma$}
\label{sec:mass}

We will use the $\mu \rightarrow e + \gamma$ branching ratio
calculation of reference~\cite{dave}, which calculates all leading
one loop contributions.  Previous analyses omitted several
significant contributions, often even the largest ones.

Neglecting the $A$ terms and the third family, the calculation
includes only three $2 \times 2$ mass matrices: the Yukawa matrix,
the lepton doublet scalar mass matrix, and the electron singlet
scalar mass matrix.  The associated physical parameters are the two
Yukawa eigenvalues; two scalar mass eigenvalues for both the lepton
doublet and electron singlet; and a mixing angle for both the lepton
doublet, $\theta_{l}$, and electron singlet, $\theta_{\bar{e}}$,
that
describes the rotation between the sparticle and particle mass
eigenbases.

Because the scalar mass splittings are required to be small, we will
parametrize the doublet and singlet scalar mass eigenvalues by the
average masses, $m_{l}^2$ and $m_{\bar{e}}^2$, and the mass
splittings, $\delta \tilde{m}_{l}^2$ and $\delta
\tilde{m}_{\bar{e}}^2$.  We will also keep only the leading
contribution in both the mass splittings and the mixing angles.
Equation~\ref{eq:gamp} gives the branching ratio for the process
$\mu
\rightarrow e+\gamma$.  The functions $X_{l}$ and $X_{\bar{e}}$ are
given in appendix~\ref{sec:functions}.
\begin{equation}
\label{eq:mue}
BR(\mu \rightarrow e+\gamma)= \frac{3e^{2}}{2\pi^{2}} \left\{
\theta_{l}^{2} \left( \frac{M_{w}}{m_{\tilde{l}}} \right)^{4}
(X_{l})
^{2} \left( \frac{\delta \tilde{m}_{l}^2}{m_{\tilde{l}}^{2}}
\right)^2 + \theta_{\bar{e}}^{2} \left(
\frac{M_{w}}{m_{\tilde{\bar{e}}}} \right)^{4} (X_{\bar{e}}) ^{2}
\left( \frac{\delta m_{\tilde{\bar{e}}}^2}{m_{\tilde{\bar{e}}}^{2}}
\right)^2 \right\}
\label{eq:gamp}
\end{equation}

\subsection{$K^{0}-\bar{K}^{0}$}

We will use the $K^{0}-\bar{K}^{0}$ mixing calculation of reference
\cite{fcnc} which computed the dominant supersymmetric contribution,
the gluino box diagrams.  Because there are no charged currents, the
weak singlet up quark does not appear in our calculation.  The
parameters in this calculation are the same as in the $\mu
\rightarrow e + \gamma$ calculation with the quark doublet and down
quark singlet replacing the lepton doublet and electron singlet.  We
will therefore use a parallel notation.  The $q$ subscript refers to
the quark doublet, and $\bar{d}$ refers to the down quark singlet.

The kaon mass splitting is given in equation~\ref{eq:kkbar}.  The
definitions of $f_{1}$ and $f_{2}$ are given in
appendix~\ref{sec:functions}.
\begin{eqnarray}
\label{eq:kk}
\Delta M_{K}= \frac{\alpha_{s}^{2}}{216 m_{\tilde{q}}^{2}} \left(
\frac{2}{3} f_{K}^{2} m_{K} \right) \left\{ \theta_{dl}^{2} \left(
\frac{\delta \tilde{m}_{\tilde{q}}^2}{m_{\tilde{q}}^2} \right)^2
f_{1} \! \left( \frac{M_{\tilde{g}}^2}{m_{\tilde{q}}^2} \right) +
\right. \nonumber \\ \left. \theta_{q}\theta_{\bar{d}} \left(
\frac{\delta m_{\tilde{q}}^2}{m_{\tilde{q}}^2} \right) \left(
\frac{\delta \tilde{m}_{\bar{d}}^2}{m_{\tilde{\bar{d}}}^2} \right)
f_{2} \! \left( \frac{M_{\tilde{g}}^2}{m_{\tilde{q}}^2} \right) +
\theta_{\bar{d}}^{2} \left( \frac{\delta
\tilde{m}_{\bar{d}}^2}{m_{\tilde{\bar{d}}}^2} \right)^2 f_{1} \!
\left( \frac{M_{\tilde{g}}^2}{m_{\tilde{\bar{d}}}^2} \right)
\right\}
\label{eq:kkbar}
\end{eqnarray}

\subsection{Experimental Constraints}

\begin{figure}
	\epsfysize 8.5cm
	\centerline{\epsffile{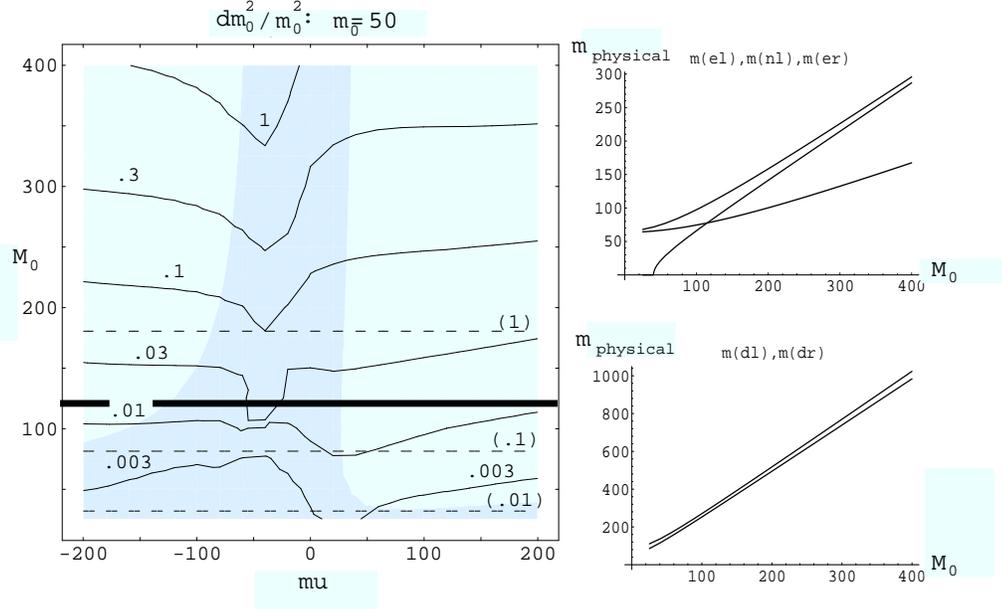}}
	\caption{$m_0=50$ GeV: Fractional mass splittings ({\it
left}) and
physical masses of sleptons ({\it top right}) and down squarks ({\it
bottom right}).}
	\label{fig:m50}
\end{figure}

\begin{figure}
	\epsfysize 8.5cm
	\centerline{\epsffile{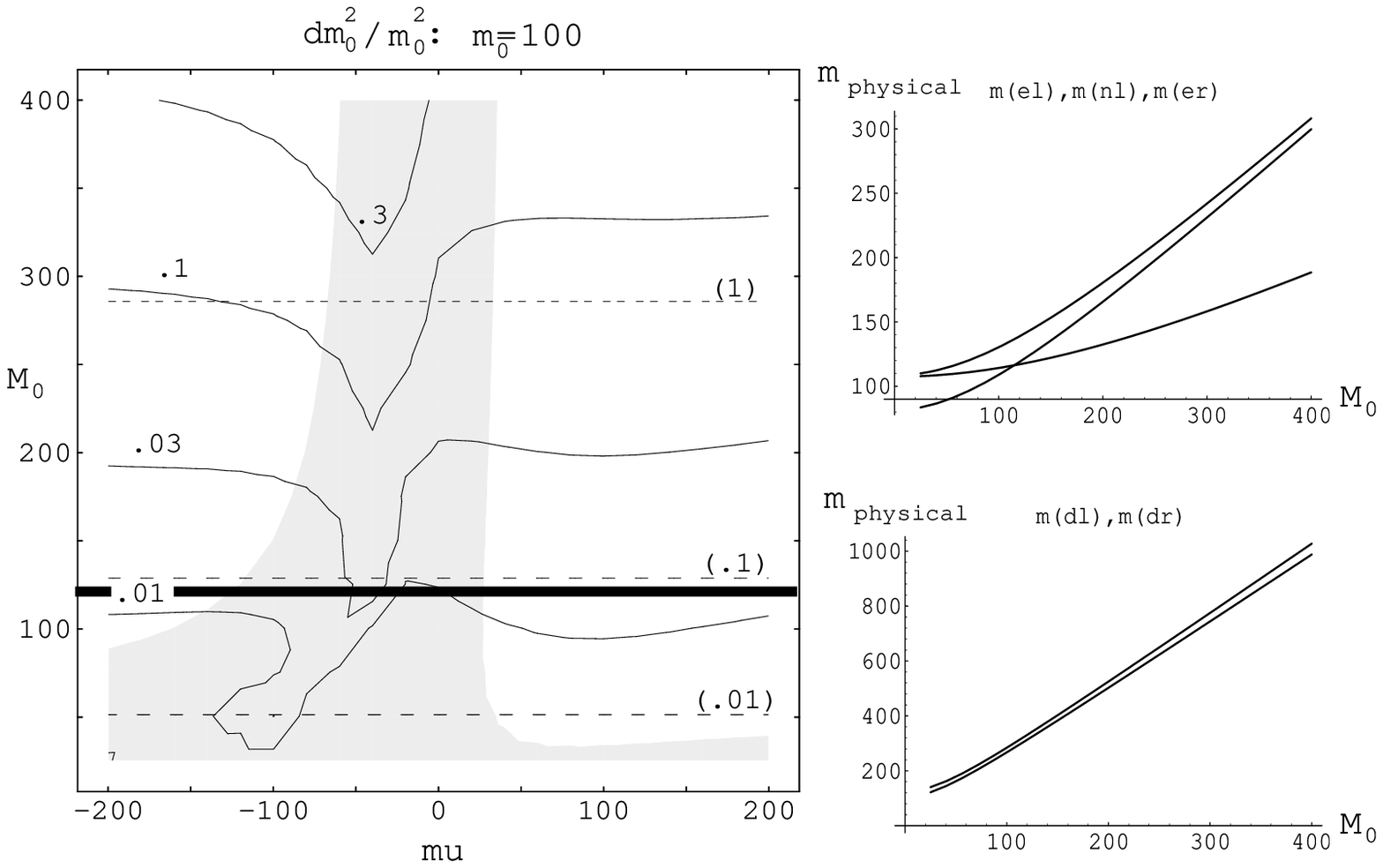}}
	\caption{$m_0=100$ GeV: Fractional mass splittings ({\it
left}) and
physical masses of sleptons ({\it top right}) and down squarks ({\it
bottom right}).}
	\label{fig:m100}
\end{figure}

\begin{figure}
	\epsfysize 8.5cm
	\centerline{\epsffile{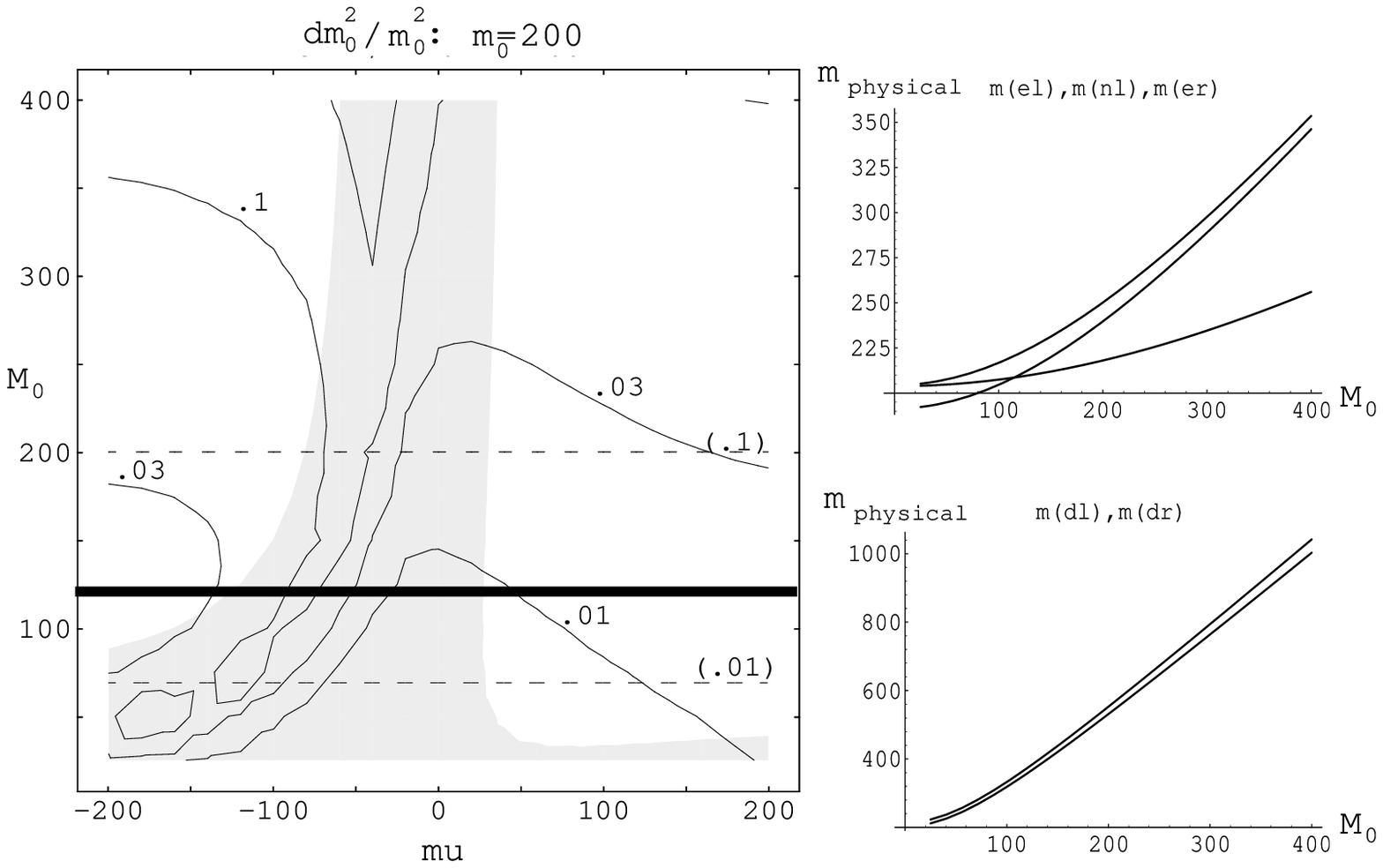}}
	\caption{$m_0=200$ GeV: Fractional mass splittings ({\it
left}) and
physical masses of sleptons ({\it top right}) and down squarks ({\it
bottom right}).}
	\label{fig:m200}
\end{figure}

\begin{figure}
	\epsfysize 8.5cm
	\centerline{\epsffile{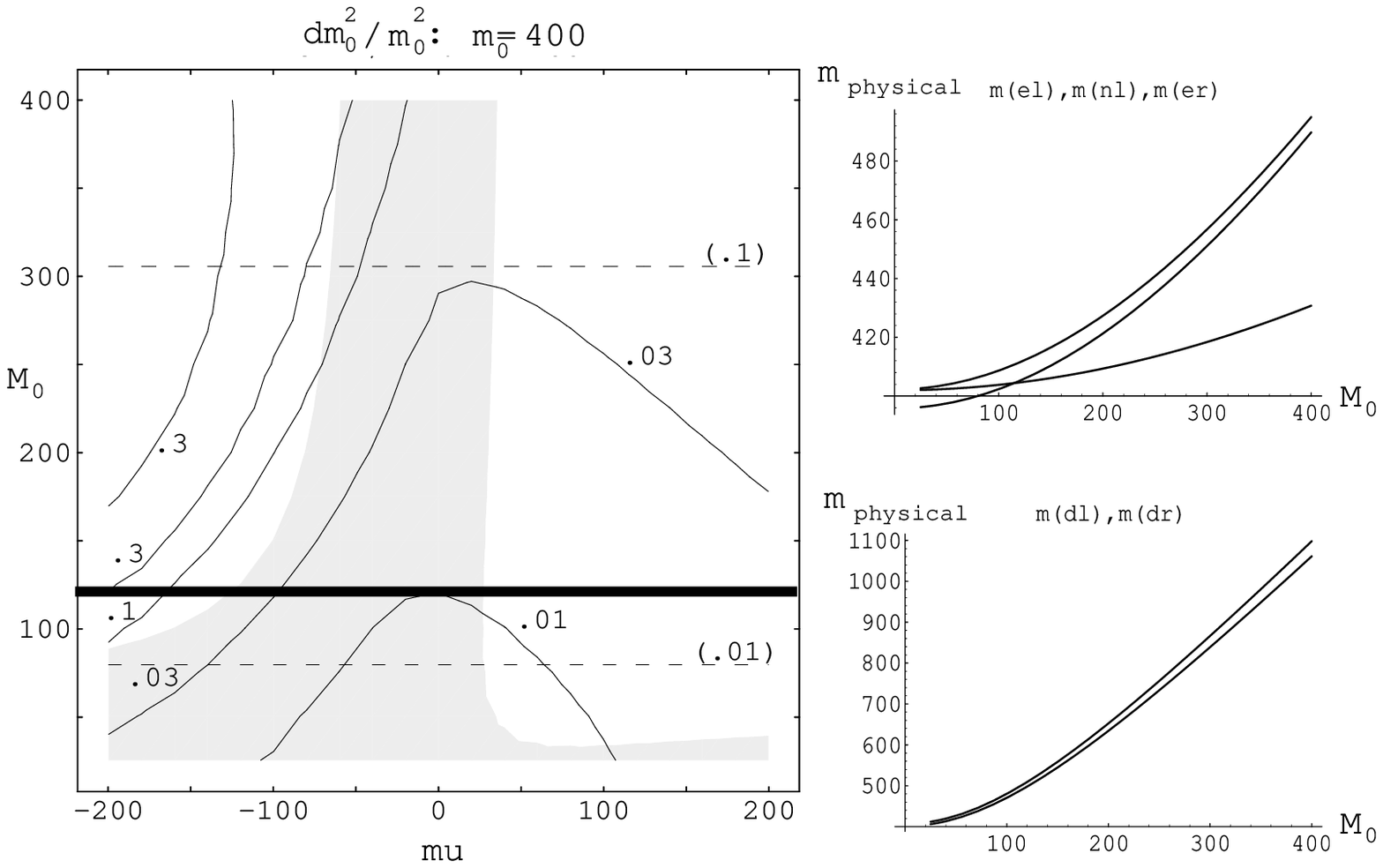}}
	\caption{$m_0=400$ GeV: Fractional mass splittings ({\it
left}) and
physical masses of sleptons ({\it top right}) and down squarks ({\it
bottom right}).}
	\label{fig:m400}
\end{figure}

We will take equations~\ref{eq:mue}~and~\ref{eq:kk} and solve them
for the fractional scalar mass splitting $\delta m^2 / m^2$.  We
then
use the one loop RG equations to relate the low energy result to the
fundamental scale, which we assume is $M_{gut}$ for the graphs of
figures~\ref{fig:m50}-\ref{fig:m400}.  Because we ignore the
contribution of the two lightest generation Yukawa couplings, the
only source of mass splitting betwen the first and second family
will
be the boundary conditions at the GUT scale.

For our graphs we choose $\tan{\beta}=3$ and, as stated above,
$A=0$.
The mass splitting constraint gets stricter for larger values of
$\tan{\beta}$, but does not change much as $\tan{\beta}$ gets
smaller.  For simplicity of presentation, we assume the singlet and
doublet mixing angles and mass splittings are the same.
Furthermore,
we assume each mixing angle is equal to the square root of the
masses
of the two particles it relates: for the leptons,
$\theta_{l}=\theta_{\bar{e}}=\sqrt{e/\mu}$, and for the quarks,
$\theta_{q}=\theta_{\bar{d}}=\sqrt{d/s}$.  The result holds, at
least
approximately, in most unified theories of fermion masses \cite{an}
and is a consequence of  quark-lepton unification and the   the
successful relation: $\theta_{\rm Cabibbo}=\sqrt{d/s}$.

Figures~\ref{fig:m50} through~\ref{fig:m400} are contour plots of
the
upper limits on the fractional scalar mass splittings, evaluated at
the GUT scale, as a function of SUSY parameter space.  We show four
graphs, one for each of four values of the scalar mass evaluated at
the GUT scale, $m_0$.  The axes of the graphs are the Higgs mixing
parameter evaluated at the weak scale, $\mu$, and the gaugino mass
evaluated at the GUT scale, $M_0$. The solid contours are the upper
limit of the fractional mass splitting of the sleptons from $\mu
\rightarrow e+\gamma$.  The dashed lines, which are labeled in
parentheses, are the upper limit of the fractional mass splitting of
the down and strange squarks from $K^{0}-\bar{K}^{0}$ mixing.  We
have also included a bold line at $M_0=120$ GeV which is the maximum
value of $M_0$ based on the 10\% naturalness criterion \cite{bg}.
The shaded region is the experimentally excluded region where the
lightest chargino is less than 45 GeV.

Accompanying each contour plot, we have included two graphs which
give the associated physical masses of the three sleptons and two
down squarks as a function of $M_0$.

These constraints can easily be adapted for new values of the mixing
angles or the branching ratio.  The resulting fractional mass
splittings may be read from the contour graphs using
equations~\ref{eq:out1} and~\ref{eq:out2}.
\begin{eqnarray}
\label{eq:out1}
\left( \frac{\delta m_0^2}{m_0^2} \right)_{lepton} & = (\frac{\delta
m_0^2}{m_0^2}{\rm~from~graph})_{\mu \rightarrow e+\gamma} \left\{
\frac{BR(\mu \rightarrow e+\gamma)}{4.9 \times 10^{-11}}
\right\}^{\frac{1}{2}} \left\{ \frac{.07}{\theta_{lepton}} \right\}
\\
\label{eq:out2}
\left( \frac{\delta m_0^2}{m_0^2} \right)_{down} & = (\frac{\delta
m_0^2}{m_0^2}{\rm~from~graph})_{K^{0} - \bar{K}^{0}} \left\{
\frac{.22}{\theta_{down}} \right\}
\end{eqnarray}

If we take the upper limit of $M_0=120$ GeV from the 10\% fine
tuning
criterion, we see that the upper limit to the slepton fractional
mass
splitting is about .01 on all the graphs.  There is an exception to
this is for large values of $m_0$ (400 GeV) and a negative value for
$\mu$.  The amplitude for the decay to a final state left handed
electron passes through zero here, leaving only the less important
right handed electron contribution and making the limit not as
strong.  However, the $K^0-\bar{K}^0$ mixing constraint is important
in this range of parameter space, and we still obtain a mass
splitting limit near .01, this time for the down-type quarks.  In
light of the mass splitting induced by a threshold correction at the
flavor scale (section 3), this mass splitting is unnaturally small.

We can obtain more reasonable mass splitting limits if we relax the
$M_0=120$ GeV constraint.  If we allow $M_0 \approx 300$ GeV or
$400$
GeV, the fractional mass splitting constraints are weakened to
$\approx$ .1 to .3.  However, this requires fine tuning of 1\% from
the naturalness criterion.  In other words, the apparently unrelated
terms in the equation for electroweak breaking,
equation~\ref{eq:mz2}, sum to give an answer 100 times smaller than
the individual terms.  This is difficult to swallow unless a
cancellation mechanism exists.

We can not simultaneously satisfy constraints from naturalness and
flavor differentiation.  This implies that there must be a mechanism
that supresses the supersymmetric contribution to flavor changing
processes.

\section{A Case for Light Messengers}

The present paper has focused on the conflict between the following
statements:
\begin{itemize}
\item[1)]
Naturalness implies light sparticles;
\item[2)]
Theories attempting to explain (even small parts of) the problem of
flavor predict large sparticle splittings;
\item[3)]Suppression of rare processes implies that sparticles are
either very heavy or highly degenerate.
\end{itemize}

One way to resolve the conflict is to decouple the physics of
fermionic flavour from that of the soft terms and thus evade
statement (2).  Consider, for example, a theory in which the soft
terms shut off above a scale $\Lambda \ll M_{\rm PL}$ (or $M_{\rm
GUT}$).  In such a theory the soft terms would not be distorted by
the flavour physics that takes place at $\sim  M_{\rm PL}$  (or
$M_{\rm GUT}$) and gives rise to the ordinary quark and lepton
masses.  If the soft terms are generated at the scale $\Lambda \ll
M_{\rm PL}$  and satisfy universality and proportionality then they
will not cause any large flavour violations near the weak scale.
The
deviations from universality and proportionality that arise between
the scales $\Lambda$ and $ M_{\rm W}$ are caused by the ordinary
Yukawa
couplings and are harmless.

\begin{figure}
\centerline{\epsffile{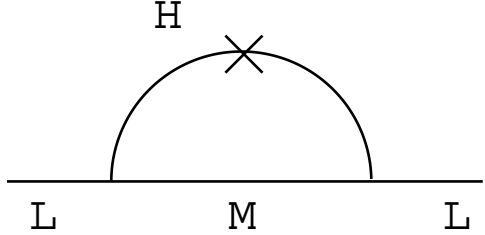}}
\caption{Schematic diagram illustrating how SUSY breaking is
communicated from the hidden sector $H$ via a messenger $M$ to $SU_3
\times SU_2 \times U_1$ carrying sparticles $L$.  $L$ can be light
$\sim M_W$ or heavy $\sim M_{\rm GUT}$.}
\label{fig:hidden}
\end{figure}

An interesting class of such  theories are those with dynamically
broken supersymmetry near the weak scale~\cite{dine}.  Another class
are (scaled down versions of) the geometric hierarchy type
theories\cite{dr}.  These are theories in which SUSY breaking
originates in a hidden sector $(H)$ and is communicated to the
particles carrying $SU_3 \times SU_2 \times U_1$ quantum numbers
$(L)$ via messengers $(M)$ as pictured in Fig.~\ref{fig:hidden}.
The
particles $L$ carry $SU_3 \times SU_2 \times U_1$ quantum numbers
and
can be light $\sim M_{\rm weak}$ or heavy $\sim M_{\rm GUT}$.

The soft masses induced by Fig.~\ref{fig:hidden} are, for example,
of
the form
\begin{equation}
\tilde m_L \simeq \frac{M^2_H}{M_M} \sim M_{\rm weak}
\label{soft}
\end{equation}
where $M_M$ is the messenger mass and $M_H$ is a SUSY breaking mass.
In the geometric hierarchy models \cite{dr}, $M_M \sim M_{\rm GUT}$
and $M_H \sim 3 \times 10^9$~GeV.  In models where the messenger is
gravity \cite{sugra}  $M_M \sim M_{\rm PL}$ and $M_H \sim 3 \times
10^{10}$~GeV. It is not difficult to consider geometric hierarchy
type models where the messenger mass $M_M$ is lighter than $M_{\rm
GUT}$ and $M_H \sim \sqrt{M_M  M_{\rm weak}}$ is proportionally
lighter.  What does one gain by this? At high momenta $p$ the soft
term $\tilde m_L$ of the above equation behaves as:
\begin{equation}
\tilde m_L(p) \simeq \frac{M^2_H}{p}
\end{equation}

It shuts off at $p \gg  M_M$, and does not feel any of the flavor
physics happening near $ M_{\rm PL}$ (or $M_{\rm GUT}$).
Consequently, the sparticle splittings and rare processes coming
from
Planckian (or GUT) physics are suppressed by powers of $M_M/M_{\rm
PL}$ (or $M_M/M_{\rm GUT}$) relative to their values in models where
the messenger is supergravity. The phenomenology of such models is
quite different from the canonical supersymmetric theories where
$M_M
\sim M_{\rm GUT}$ or $M_{\rm PL}$.  In particular, if $M_M\ll M_{\rm
GUT}$, the sparticle masses are more degenerate and deviations from
universality or proportionality are smaller.

\section{Conclusion}

The minimal supersymmetric standard model with universality and
proportionality provides a means of  preserving the light weak scale
and it is consistent with the observed flavor changing data.
However, it does not seem to arise from a fundamental theory
explaining flavor.  Models which do explain the fermionic flavor
hierarchy typically contain flavor dependence in the sparticle
masses which contributes to flavor changing interactions. The same
is
true  in any theory that contains Planck or GUT- mass particles
that couple asymmetrically to the families with strength comparable
to the gauge couplings. One
solution to this problem is to increase the mass of the sparticles.
However, this implies severe fine tuning of the parameters of the
theory.

In this paper, we have reviewed the origin of flavor dependence in
the scalar mass matrices and the fine tuning constraints
(naturalness
criterion).  We then graphed the constraints from flavor changing
processses for a general class of theories to quantify the conflict
that exists between the physics of flavor and naturalness in
electroweak breaking.  We have found that it is not possible to
simultaneously satisfy the constraints from both flavor and
naturalness, suggesting that there is some mechanism which accounts
for the observed smallness in flavor changing processes.  One
possibility is to have supersymmetry breaking transmitted to the
observed particles by messengers much lighter than the Planck scale.

\section*{Acknowledgments}
It is a pleasure to thank Uri Sarid for many valuable discussions
and
for reading the manuscript.  D.S. would like to thank the CERN
Theory
Group for its hospitality.

\appendix

\section{Special Function Definitions}
\label{sec:functions}

\subsection{$\mu \rightarrow e + \gamma$ Loop Functions}

\paragraph\ Equations~\ref{eq:loop1}~through~\ref{eq:loop4} give the
loop functions for $\mu \rightarrow e + \gamma$.  Complete details
of
the calculation are in reference~\cite{dave}.
\begin{eqnarray}
\label{eq:loop1}
f(r) & = & \frac{1}{12(1-r)^4} \left( 2r^3+3r^2-6r+1-6r^2\log{r}
\right) \\
g(r) & = & \frac{1}{12(1-r)^4} \left( r^3-6r^2+3r+2+6r\log{r}
\right)
\\
h(r) & = & \frac{1}{2(1-r)^3} \left( -r^2+1+2r\log{r} \right) \\
j(r) & = & \frac{1}{2(1-r)^3} \left( r^2-4r+3+2\log{r} \right)
\label{eq:loop4}
\end{eqnarray}

\subsection{$\mu \rightarrow e + \gamma$ Amplitude Functions}

In section~\ref{sec:mass}, we calculated the transition amplitude in
the context of a particular theory of lepton masses.  In this
appendix we give the necessary functions for equation~\ref{eq:gamp}.
We use modified loop functions which are defined below.  The
argument
of these loop function is $r_{pk}=M_{k}^{2}/m_{p}^2$ where $k$
represents the chargino or neutralino, and $p$ represents the
slepton.

The $U$ matrices rotate the gaugino/higgsino interaction basis into
the neutralino/chargino mass basis.  $U^{0}$ is for the neutralinos;
$U^{+}$ is for the charginos $\tilde{W}^{+}$ and
$\tilde{H}_{u}^{+}$;
and $U^{-}$ is for the charginos $\tilde{W}^{-}$ and
$\tilde{H}_{d}^{-}$.
\begin{eqnarray}
X_{l} & = & X_{lf} + X_{lh} + X_{lg} + X_{lj} \\
X_{r} & = & X_{rf} + X_{rh} \\
X_{lf} & = & \frac{1}{2} \left( U^{0}_{Wk} + \frac{g_{1}}{g_{2}}
U^{0}_{Bk} \right)^{2} f_{g}(r_{ek}) \\
X_{rf} & = & -2 \left( \frac{g_{1}}{g_{2}} U^{0}_{Bk} \right)^{2}
f_{g}(r_{\bar{e}k}) \\
X_{lh} & = & \frac{(A+\mu \tan{\beta}) M}{m^{2}_{\tilde{e}}} \left(
\frac{g_{1}}{g_{2}} U^{0}_{Bk} \right) \left( U^{0}_{Wk} +
\frac{g_{1}}{g_{2}} U^{0}_{Bk} \right) h_{k}(r_{ek},r_{\bar{e}k})
\nonumber \\
& & - \frac{M}{\sqrt{2} g_{2} v_{1}} \left( U^{0}_{H k} \right)
\left( U^{0}_{Wk} + \frac{g_{1}}{g_{2}} U^{0}_{Bk} \right)
h_{g}(r_{ek}) \nonumber \\
& & + \left( \frac{m^{4}_{\tilde{e}}}{\delta m^{2}_{\tilde{e}}}
\right) \frac{\delta A_{\bar{\mu} e}
M}{m^{2}_{\tilde{e}}-m^{2}_{\tilde{\bar{e}}}} \left(
\frac{g_{1}}{g_{2}} U^{0}_{Bk} \right) \left( U^{0}_{Wk} +
\frac{g_{1}}{g_{2}} U^{0}_{Bk} \right) \left[
\frac{h(r_{ek})}{m^{2}_{\tilde{e}}} -
\frac{h(r_{\bar{e}k})}{m^{2}_{\tilde{\bar{e}}}} \right] \\
X_{rh} & = & \frac{(A+\mu \tan{\beta}) M}{m^{2}_{\tilde{\bar{e}}}}
\left( U^{0}_{Wk} + \frac{g_{1}}{g_{2}} U^{0}_{Bk} \right) \left(
\frac{g_{1}}{g_{2}} U^{0}_{Bk} \right) h_{k}(r_{\bar{e}k},r_{ek})
\nonumber \\
& & + \frac{\sqrt{2} M}{g_{2} v_{1}} \left( U^{0}_{H k} \right)
\left( \frac{g_{1}}{g_{2}} U^{0}_{Bk} \right)^{2}
h_{g}(r_{\bar{e}k})
\nonumber \\
& & + \left( \frac{m^{4}_{\tilde{\bar{e}}}}{\delta
m^{2}_{\tilde{\bar{e}}}} \right) \frac{\delta A_{\mu \bar{e}}
M}{m^{2}_{\tilde{\bar{e}}} - m^{2}_{\tilde{\bar{e}}}} \left(
U^{0}_{Wk} + \frac{g_{1}}{g_{2}} U^{0}_{Bk} \right) \left(
\frac{g_{1}}{g_{2}} U^{0}_{Bk} \right) \left[
\frac{h(r_{\bar{e}k})}{m^{2}_{\tilde{\bar{e}}}} -
\frac{h(r_{ek})}{m^{2}_{\tilde{e}}} \right] \\
X_{lg} & = & -\left( \frac{m^{4}_{\tilde{e}}}{m^{4}_{\tilde{\nu}}}
\right) \left( U^{+}_{Wk} \right)^{2} g_{g}(r_{\bar{\nu}k}) \\
X_{lj} & = & \left( \frac{m^{4}_{\tilde{e}}}{m^{4}_{\tilde{\nu}}}
\right) \frac{M}{g_{2} v_{1}} \left( U^{-}_{H k} \right) \left(
U^{+}_{Wk} \right) j_{g}(r_{\bar{e}k})
\end{eqnarray}

Although we set $A=0$ in the main text, we include the $A$
dependence
here in the appendix.  Equations~\ref{eq:a1}~and~\ref{eq:a2} give
definitions for the nonuniversality of the $A$ terms used above.
\begin{eqnarray}
\label{eq:a1}
\delta A_{\bar{\mu} e} = A_{\bar{\mu} e} - A_{\bar{\mu} \mu} \\
\delta A_{\bar{e} \mu} = A_{\bar{e} \mu} - A_{\bar{\mu} \mu}
\label{eq:a2}
\end{eqnarray}

For our original functions $f$, $g$, $h$, and $j$, we have two
modifications that result from our expansion in the inter-family
mass
difference.  Equation~\ref{eq:gim} defines the $g$ subscript, and
equation~\ref{eq:kim} defines the $k$ subscript.  $Z$ represents any
of the four functions $f$, $g$, $h$, or $j$.
\begin{eqnarray}
\label{eq:gim}
Z_{g} \left( \frac{M^{2}}{m^{2}} \right) & \equiv & m^4
\frac{d}{dm^2} \left\{ \frac{1}{m^2} Z \left( \frac{M^2}{m^2}
\right)
\right\} \\
Z_{k} \left( \frac{M^{2}}{m^{2}_{a}} , \frac{M^{2}}{m^{2}_{b}}
\right) & \equiv & m^6_{a} \frac{d}{dm^2_{a}} \left\{
\frac{1}{m^{2}_{a} - m^{2}_{b}} \left[ \frac{1}{m^{2}_{a}} Z \left(
\frac{M^2}{m^{2}_{a}} \right) - \frac{1}{m^{2}_{b}} Z \left(
\frac{M^2}{m^{2}_{b}} \right) \right] \right\}
\label{eq:kim}
\end{eqnarray}

\subsection{$K^{0}-\bar{K}^{0}$ Functions}

\paragraph\  Equations~\ref{eq:kk1}~and~\ref{eq:kk2} are the loop
functions from the text in terms of the functions $f_{6}$ and
$\tilde{f_{6}}$ of Hagelin et. al., shown in
equations~\ref{eq:hag1}~and~\ref{eq:hag2}.

\begin{eqnarray}
\label{eq:kk1}
f_{1}(r) & = & -66 \tilde{f_{6}}(r) - 24 r f_{6}(r) \\
\label{eq:kk2}
f_{2}(r) & = & \left\{ -36-24 \left( \frac{m_{K}}{M_{s}+m_{d}}
\right) ^2 \right\} \tilde{f_{6}}(r) + \nonumber
\\       &   & \left\{ -72+384 \left( \frac{m_{K}}{m_{s}+m_{d}}
\right) ^2 \right\} r f_{6}(r) \\
\label{eq:hag1}
f_{6}(r) & = & \frac{1}{6(1-r)^5} \left(
-r^3+9r^2+9r-17-18r\log{r}-6r\log{r} \right) \\
\label{eq:hag2}
\tilde{f_{6}}(r) & = & \frac{1}{3(1-r)^5} \left(
r^3+9r^2-9r-1-6r^2\log{r}-6r\log{r} \right)
\end{eqnarray}

\end{document}